\documentstyle[12pt,epsf,epsfig]{article}
\begin{document}

\title{X(1835): A Possible Baryonium?}
\author{Shi-Lin Zhu$^{1,2}$ And Chong-Shou Gao$^1$ \\
$^1$Department of Physics, Peking University, Beijing 100871,
China\\
$^2$RCNP, Osaka University, Japan} \maketitle

\begin{abstract}

We point out that (1) the large $p\bar p$ coupling and suppressed
mesonic coupling of X(1835) and (2) the suppression of the
three-body strange final states strongly indicate that X(1835) may
be a $p\bar p$ baryonium. We also point out that the branching
ratio of $X(1835)\to\eta \pi\pi$ should be bigger than that of
$X(1835)\to\eta^\prime \pi\pi$. If BES further confirms the
non-observation of X(1835) in the $\eta\pi\pi$ channel, that will
be very puzzling. Finally, X(1835) may be used a tetraquark
generator if X(1835) is really established as a baryonium state.

\end{abstract}
\vspace{0.3cm}

\pagenumbering{arabic}

\section{Introduction}

BES Collaboration observed a significant threshold enhancement of
$p\bar p$ mass spectrum in the radiative decay $J/\psi\to \gamma
p\bar p$ \cite{bes}. No similar signal was observed in the channel
$\pi^0 p\bar p$. Ignoring the final state interaction, the central
value of this enhancement from S-wave fit was around 1859 MeV
\cite{bes}. With the final state interaction in the isoscalar
channel calculated in Ref. \cite{fsi} as input, BES refit the mass
and found it lies around 1830 MeV with a width $\Gamma =(0\pm 93)$
MeV \cite{jin}.

Theoretical work speculated many possibilities for the enhancement
such as the t-channel pion exchange, some kind of threshold
kinematical effects, a new resonance below threshold or even a
$p\bar p$ bound state etc
\cite{yan,q1,q2,q3,q4,q5,q6,q7,q8,q9,q10,q11,q12,rosner}.

In order to establish this enhancement as a new resonance X(1835),
we argued that this state must be observed in the mesonic final
states \cite{gao}. Based on the non-observation of X(1835) in the
following final states $\pi^+\pi^-, 2\pi^0, K{\bar K}, 3\pi$ at
that time \cite{sxy}, we concluded that the quantum number of this
possible signal is very likely to be $J^{PC}=0^{-+}, I^G=0^+$
\cite{gao}. We called for the experimental search of X(1835) in
the mesonic decay channels according to their priority: $\eta
\pi\pi$, $\eta'\pi\pi$, $\eta\eta\eta$, $4\pi$, $K{\bar K}\pi$,
$\eta K{\bar K}$, $K\bar K \pi\pi$, 6$\pi$ \cite{gao}.

Last month BES Collaboration reported preliminary results on
X(1835) in the $J/\psi \to \gamma \eta^\prime \pi^+\pi^-$ channel
\cite{jin}. The $\eta^\prime$ meson was detected in both
$\eta\pi\pi$ and $\gamma\rho$ channels. There are roughly $264\pm
54$ events. With a statistical significance of $7.7\sigma$, BES
Collaboration measured the mass of X(1835) to be $(1833.7\pm
6.2\pm 2.7)$ MeV and its width to be $(67.7\pm 20.3\pm 7.7)$ MeV
\cite{jin}. To the present authors, BES's observation of X(1835)
in the $\eta^\prime \pi\pi$ is very encouraging while its
non-observation of X(1835) in the $\eta \pi\pi$ is a little
confusing. In this short note, we shall try to explore the
underlying structure of X(1860) using the available experimental
information.

\section{Expected typical properties of a baryonium}

Because of its large $p\bar p$ branching ratio and close proximity
to $p\bar p$ threshold, the possibility of X(1835) being a
baryonium is particularly interesting. The study of nucleon and
anti-nucleon bound states dated back to Fermi and Yang
\cite{fermi}. An extensive theoretical and experimental review can
be found in Refs. \cite{history,pr}.

In the sector of nucleon-nucleon interaction, the short-range part
was constrained and well determined using the properties of the
light nuclei like deuteron and the available enormous
nucleon-nucleon scattering data as inputs. In contrast, the
nucleon anti-nucleon scattering data is scarce. The short-range
part of $N\bar N$ interaction $V_{N\bar N}$ remains essentially
unknown. Especially the annihilation contribution is very
difficult to take into account. Because of very poor knowledge of
the short range part of $V_{N\bar N}$, some deeply-bound $N\bar N$
states are always predicted using the phenomenological $N\bar N$
potential. Experimentally none of them was found. A reliable
calculation of the spectrum of $N\bar N$ bound states is still too
demanding at present.

In the following, we list some typical properties of X(1835) as a
baryonium which one would naively expect.
\subsection{Very large coupling of X(1835) with $p\bar p$}

According to BES's measurement, $BR(J/\psi \to \gamma X)\cdot
BR(X\to p\bar p) \sim 7\times 10^{-5}$ \cite{jin}. Using
$BR(J/\psi \to \gamma X)\sim (0.5-2)\times 10^{-3}$ for a $0^{-+}$
meson, $BR(X\to p\bar p)\sim (4-14)\%$ assuming $\Gamma_X <30$ MeV
\cite{bes}. This branching ratio will increase if $\Gamma_X$
increases \cite{jin}. Recall that $2m_p=1876$ MeV while $m_X=1835$
MeV and $\Gamma_X\approx 68$ MeV. The decay of X(1835) into $p\bar
p$ occurs only from the tiny tail of its mass distribution. We may
write an effective Lagrangian for this process:
\begin{equation}
{\cal L}=g_{Xp\bar p} {\bar p}i\gamma_5 p X + \mbox{H.c.} \; .
\end{equation}
Then the decay width reads
\begin{equation}\label{breit}
\Gamma (X\to p\bar p) ={1\over 8\pi}|g_{Xp\bar p}|^2
\int_{4m_p^2}^{(M_X+\Gamma_X)^2} ds \sqrt{s-4 m_p^2} f(s, M_X,
\Gamma_X)
\end{equation}
where
\begin{equation}
f(s, M_X, \Gamma_X)={1\over \pi} {M_X\Gamma_X \over (s-M_X^2)^2
+M_X^2 \Gamma_X^2}
\end{equation}
is the Breit-Wigner distribution. Note we have taken the upper
limit of the integral to be $(M_X+\Gamma_X)^2$ instead of
infinity. The reason is simple. The Breit-Wigner distribution is
valid only for very narrow resonances. For any realistic case, the
narrow resonance distribution should die off very quickly at
$\sqrt{s}=M_X +\Gamma_X$. Numerically we have
\begin{equation}
%\Gamma (X\to p\bar p) =1.8\times |g_{Xp\bar p}|^2 MeV for M_X+2\Gamma_X
\Gamma (X\to p\bar p) =0.57 \times |g_{Xp\bar p}|^2 \mbox{MeV}
\end{equation}
Even with $BR(X\to p\bar p)\sim 10\%$, we get $g_{X p\bar
p}\approx 3.5$ or
\begin{equation}\label{11}
\alpha_{X p\bar p}={g_{X p\bar p}^2\over 4\pi} \approx 1.0 \;.
\end{equation}

Now let's move on to the decay mode $X(1835)\to \eta^\prime
\pi^+\pi^-$. Its decay width can be estimated using the following
formula:
\begin{equation}
\Gamma(X\to \eta^\prime \pi^+\pi^-) \approx \alpha_{X \eta^\prime
\pi\pi}{{\bar k}^{2L+1}\over M_X^{2L}}
\end{equation}
where $\bar k\approx 400$ MeV is the averaged decay momentum and
$L$ is the decay angular momentum. The above decay occurs through
S-wave. Hence L=0. Assuming $\Gamma(X\to \eta^\prime
\pi^+\pi^-)\approx 20$ MeV, we arrive at
\begin{equation}
\alpha_{X \eta^\prime \pi\pi}\approx 0.05
\end{equation}
which is in strong contrast with Eq. (\ref{11}).

The strong enhancement of $p\bar p$ coupling and severe
suppression of mesonic coupling indicates that there may exist a
large $p\bar p$ component in the wave function of X(1860).
Invoking the baryonium hypothesis, the coupling difference can be
ascribed to the underlying structure of X(1835) easily.

\subsection{Suppression of strange three-body final states}

As a $p\bar p$ bound state, X(1835) contains three up and down
quark anti-quark pairs. X(1835) can fall apart into three pairs of
non-strange mesons through color recombination. Since there is no
valence strange quark within the proton, the three-body decays
$X(1835)\to K\bar K \pi$ etc are suppressed by OZI rule. This
simple conclusion is consistent with BES's non-observation of
X(1835) in the $K\bar K\pi$ channel and observation in the
$\pi^+\pi^-\eta^\prime$. However, the four or five-body strange
decay channels are not suppressed compared with those non-strange
four/five body channels.

\section{Puzzling decay pattern}

In the chiral limit $m_q\to 0$, there are eight massless Goldstone
bosons in QCD. Because of the axial anomaly, the $SU(3)$ flavor
singlet pseudoscalar meson $\eta_1$ obtains finite mass even in
the chiral limit. In fact, the majority of the $\eta_1$ mass comes
from the anomaly. However, if we further take the large $N_c$
limit $N_c\to \infty$, $\eta_1$ becomes massless \cite{dynamics}.
Now we have a massless nonet M:
\begin{eqnarray}
M=\left(
\begin{array}{ccc}
{\pi^0\over \sqrt{2}}+{\eta_8\over \sqrt{6}}+{\eta_1\over \sqrt{3}}&\pi^+ &K^+\\
\pi^-&-{\pi^0\over \sqrt{2}}+{\eta_8\over \sqrt{6}}+{\eta_1\over \sqrt{3}}&K^0\\
K^-&{\bar K}^0&-{2\over \sqrt{6}}\eta_8+{\eta_1\over \sqrt{3}}
\end{array}
\right).
\end{eqnarray}
We may write down the effective $p\bar p$ three pseudoscalar meson
interaction Lagrangian.
\begin{equation}
{\cal L}=g_D \mbox{Tr} \left( \bar B \{M^3, B\}_+\right) +g_F
\mbox{Tr} \left( \bar B [M^3, B]_-\right)
\end{equation}
where the matrix $B$ is the nucleon octet:
\begin{eqnarray}
B=\left(
\begin{array}{ccc}
{\Sigma^0\over \sqrt{2}}+{\Lambda\over \sqrt{6}}& \Sigma^+&p\\
\Sigma^-&-{\Sigma^0\over \sqrt{2}}+{\Lambda\over \sqrt{6}}&n\\
\Xi^-&\Xi^0&-{2\over \sqrt{6}}\Lambda
\end{array}
\right).
\end{eqnarray}
After extracting the $p\bar p$ term, we have
\begin{equation}\label{22}
{\cal L}\sim X(1860)\cdot \left( {\pi^0\pi^0\over
2}+\pi^+\pi^-\right)\cdot \left(\eta_1 +{\eta_8\over
\sqrt{2}}\right)
\end{equation}
where we have replaced $p\bar p$ by X(1835). Naively one finds the
coupling between X(1835) and $\eta_1\pi\pi$ is a factor of
$\sqrt{2}$ larger than that between X(1835) and $\eta_8\pi\pi$.
However the physical states are $\eta, \eta^\prime$, which is a
mixture of $\eta_1, \eta_8$:
\begin{eqnarray}\nonumber
|\eta\rangle=\cos\theta |\eta_8\rangle -\sin\theta |\eta_1\rangle
\;,\\
|\eta^\prime\rangle=\sin\theta |\eta_8\rangle +\cos\theta
|\eta_1\rangle
\end{eqnarray}
with the mixing angle $\theta\approx -{\pi\over 9}$ \cite{pdg}.
After inserting the above expressions into Eq. (\ref{22}) we have
\begin{equation}
{\cal L}\sim X(1860)\cdot \left( {\pi^0\pi^0\over
2}+\pi^+\pi^-\right)\cdot \left(0.7 \eta^\prime +1.0 \eta_8
\right).
\end{equation}
From the above equation, it's clear that (1) the decay ratio of
$X(1835)\to\eta^\prime \pi^0\pi^0$ mode is only a factor of four
smaller than $\eta^\prime \pi^+\pi^-$ mode. BES may be able to
measure it; (2) $\eta\pi\pi$ modes are a factor of two bigger than
$\eta^\prime \pi\pi$ modes even if we ignore the larger phase
space. Therefore, BES's non-observation of $\eta \pi\pi$ modes is
really puzzling.

We can interpret the above results from a more intuitive and
transparent point of view. Let's decompose the quark content of
$\eta, \eta^\prime$ explicitly.
\begin{eqnarray}\nonumber
\eta \approx {\cos\theta\over \sqrt{6}} \left(u\bar u+d\bar
d-2s\bar s\right)-{\sin\theta\over \sqrt{3}} \left(u\bar u+d\bar
d+s\bar s\right)=0.58\left(u\bar u+d\bar d\right)-0.57s\bar s\;
,\\
\eta^\prime \approx {\sin\theta\over \sqrt{6}} \left(u\bar u+d\bar
d-2s\bar s\right)+{\cos\theta\over \sqrt{3}} \left(u\bar u+d\bar
d+s\bar s\right)=0.40\left(u\bar u+d\bar d\right)+0.82s\bar s\; .
\end{eqnarray}
If X(1835) is a $p\bar p$ bound state, it mainly couples to the
$\left(u\bar u+d\bar d\right)$ component. It's straightforward to
conclude X(1835) decays into $\eta\pi\pi$ more easily. This is in
conflict with BES's preliminary results.

One natural scapegoat to explain the inconsistency is the axial
anomaly intrinsically associated with $\eta_1$ since there may
exists lots of glue inside X(1835). Let's take a closer look at
the $J/\psi$ radiative decay which is generally believed to be
glue-rich. From PDG the branching ratio of $J/\psi \to
\gamma\eta^\prime$ is large: $(4.31\pm 0.30)\times 10^{-3}$ while
$J/\psi \to \gamma\eta$ was not observed \cite{pdg}. This is an
expected result since $\eta$ meson is mainly an octet which does
not couple to $G\tilde G$ after $J/\psi$ decays into $\gamma
G\tilde G$. Let's move on to the exactly same three pseudoscalar
meson radiative decays associated with X(1835). The branching
ratio of $J/\psi \to \gamma\eta \pi\pi$ is huge: $(6.1\pm
1.0)\times 10^{-3}$ while the $J/\psi \to \gamma\eta^\prime
\pi\pi$ mode can not be found from the summary table of PDG
\cite{pdg}. This counter-example shows that the axial anomaly may
not play a very significant role in the $J/\psi$
three-pseudoscalar-meson radiative decays or
three-pseudoscalar-meson decays of X(1835). Then comes the decay
pattern puzzle of X(1835): which mechanism is responsible for the
suppression of the $X(1835)\to\eta \pi\pi$ mode and enhancement of
$X(1835)\to\eta^\prime \pi\pi$ mode?

Finally, let's make one comment. Even in the extreme case that
this decay is completely dominated by the axial anomaly, the
$\eta_8$ term is absent in Eq. (\ref{22}). The mixing can still
induce $X(1835)\to \eta\pi\pi$ decays. The branching ratio between
$\eta\pi\pi$ and $\eta^\prime \pi^+\pi^-$ modes should be larger
than $\tan^2\theta =13\%$.

\section{Discussion}

From the available BES's preliminary data, we point out that the
effective coupling constant between X(1835) and $p\bar p$
$\alpha_{X p\bar p}$ is order unity while its mesonic coupling
constant $\alpha_{X \eta^\prime \pi\pi}$ is strongly suppressed.
Moreover the three-body strange final states are suppressed. All
these information points toward the possibility of X(1835) being a
$p\bar p$ baryonium.

Based on symmetry consideration and comparison with previous data,
we point out that the branching ratio of $X(1835)\to\eta \pi\pi$
should be bigger than that of $X(1835)\to\eta^\prime \pi\pi$. If
BES further confirms the non-observation of X(1835) in the
$\eta\pi\pi$ channel, that will be very puzzling.

Finally we want to point out that X(1835) can be used as a
tetraquark generator if it is really established a baryonium. A
baryonium can easily transform into a tetraquark after emitting a
light meson. In fact, BES Collaboration may search the decay final
states of X(1835) to find out whether there exists possible
tetraquark signals in the $\eta^\prime\pi$ invariant mass
spectrum.

\section{Acknowledgments}

C.S.G was supported by National Natural Science Foundation of
China under Grant No. 90103019, the Doctoral Program Foundation of
Institution of Higher Education, and the State Education
Commission of China under Grant No. 2000000147. S.L.Z. was
supported by the National Natural Science Foundation of China
under Grants 10375003 and 10421003, Ministry of Education of
China, FANEDD, Key Grant Project of Chinese Ministry of Education
(NO 305001) and SRF for ROCS, SEM.

%---------------------------------------------------------------------------

\end{document}